\documentclass[twocolumn,prl]{revtex4-1}
\usepackage[latin9]{inputenc}
\usepackage{color}
\usepackage{bm}
\usepackage{amsmath}
\usepackage{amssymb}
\usepackage{graphicx}

\makeatletter

\usepackage{bm}

\makeatother

\begin{document}

\title{The Haldane model under quenched disorder}

\author{Miguel Gonçalves$^{1}$, Pedro Ribeiro$^{1,2}$, Eduardo V. Castro$^{1,2,3}$}

\affiliation{$^{1}$CeFEMA, Instituto Superior Técnico, Universidade de Lisboa,
Av. Rovisco Pais, 1049-001 Lisboa, Portugal}

\affiliation{$^{2}$Beijing Computational Science Research Center, Beijing 100084,
China}

\affiliation{$^{3}$Centro de F\'{\i}sica das Universidades do Minho e Porto,
Departamento de F\'{\i}sica e Astronomia, Faculdade de Ciências, Universidade
do Porto, 4169-007 Porto, Portugal}
\begin{abstract}
We study the half-filled Haldane model with Anderson and binary disorder and
determine its phase diagram, as a function of the Haldane flux and
staggered sub-lattice potential, for increasing disorder strength.
We establish that disorder stabilizes topologically nontrivial phases
in regions of the phase diagram where the clean limit is topologically
trivial. At small disorder strength, our results agree with analytical
predictions obtained using a first order self-consistent Born approximation,
and extend to the intermediate and large disorder values where this
perturbative approach fails. We further characterize the phases according
to their gapless or gapped nature by determining the spectral weight
at the Fermi level. We find that gapless topological nontrivial phases
are supported for both Anderson and binary disorder. In the binary
case, we find a reentrant topological phase where, starting from a
trivial region, a topological transition occurs for increasing staggered
potential $\eta$, followed by a second topological transition to
the trivial phase for higher values of $\eta$. 
\end{abstract}
\maketitle

\section{Introduction}

Topological band insulators have been attracting a great deal of attention
due to their unusual properties when compared to trivial, common band
insulators \citep{RevModPhys.82.3045,QZrmp11,bernevigBook,Chiu2016}.
After the discovery of the quantum Hall effect \citep{Klitzing1980}
and its theoretical explanation using concepts of topology \citep{TKNN82,NTW85},
the first proposal of a topological phase without the need of an applied
homogeneous magnetic field was due to Haldane \citep{Haldane1988}.
The quantum anomalous Hall effect predicted by Haldane was recently
realized experimentally on several systems under zero net magnetic
field \citep{Chang2013,Jotzu2014,Checkelsky2014,Chang2015}. This
makes quantum anomalous Hall insulators, or Chern insulators according
to the topological classification \citep{Chiu2016}, very interesting
systems to study quantum topological matter at the fundamental level,
and to explore possible technological applications of their distinctive
feature -- the protected, gapless surface states that run on the
sample edges.

Topological systems are known to be robust to disorder effects \citep{Xiao2010},
meaning that the system's topological properties can emerge even in
the presence of disorder as long as it does not break any fundamental
symmetry. In the case of quantum Hall systems and anomalous quantum
Hall insulators, disorder is even crucial for the observation of quantized
Hall conductivity, since it localizes every state except for those
carrying the topological invariant \citep{KM93,OMN+03,nagaosaQSHloc07}.
A well quantized Hall plateau is then the consequence of the Fermi
level laying in the gap (filled of localized states) which separates
the extended states that carry the topological invariant. With increasing
disorder strength a localization transition occurs, accompanied by
a topological transition between a topological (Chern) insulator and
a trivial (Anderson) insulator. As disorder increases, the bulk extended
states above and below the Fermi level carrying the topological invariant
shift toward one another and annihilate, leading to the topological
phase transition into the trivial phase. The standard mechanism is
referred to as ``levitation and annihilation'' of extended states
\citep{nagaosaQSHloc07}. Increasing disorder strength leads, generally,
the Chern insulator to a trivial phase \citep{prodanBernevig,Prodan2011},
although exceptions have recently been identified \citep{Castro2015,Castro2016}.

In 2009, Li et.al. surprisingly discovered that it was possible to
obtain a topological phase transition from a topologically trivial
phase to a topologically non-trivial phase with quantized conductance
by increasing Anderson disorder \citep{Shen2009} - the now called
topological Anderson insulator (TAI) phase. This phenomenon was latter
explained by Groth et. al. with the usage of the a perturbative first
order self-consistent Born approximation \citep{Groth2009}. The picture
emerging from this low energy approach is that the trivial and non-trivial
masses compete and the phenomenon can be understood as a renormalization
of the trivial mass to lower values for increasing (although perturbative)
disorder. Building on these seminal works several studies treated
models supporting possible TAI phases, including the disordered Kane-Mele
model \citep{KM05,Orth2016} and the disordered Haldane model \citep{Song2012,Garcia2015}.
For the Kane-Mele model, the presence of a staggered potential was
shown to be a necessary condition for the disorder-driven transition
into the TAI phase and the energy window associated with a quantized
non-zero conductance was verified to increase with the disorder strength
in this case \citep{Orth2016}. In Ref.\,\citep{Song2012}, the Hall
conductance was studied in the Anderson disorder - Fermi energy plane
for the Haldane model and a TAI phase was found for a value of the
staggered potential for which the topological phase is suppressed
in the clean limit. In parallel with the disordered Kane-Mele model,
it was also observed a disorder-driven increase of the energy window
associated with the topological phase in the disordered Haldane model,
in Ref.\,\citep{Garcia2015}. The first order Born approximation
was also applied to the Haldane model in this work, by mapping the
corresponding low energy Hamiltonian into the low energy Hamiltonian
of a HgTe quantum well. Other models with similar phenomenology include
a two-dimensional semi-Dirac material \citep{Sriluckshmy2018} and
a model where the disorder is introduced through magnetic impurities
\citep{Raymond2015}. 

In the present paper we study the evolution of the total phase diagram
of the Haldane model at half-filling with increasing disorder. Despite
the knowledge accumulated on this model so far, a global phase diagram
with indicating the gapped or gapless nature of the system at each
point and its topological properties, is still lacking. Furthermore, we
extend our study of the phase diagram to both Anderson and binary
disorder, and show that interesting qualitative differences exist
between them. Our aim is to understand how topological phases behave,
not only for small disorder strength, for which topological properties
are expected to be robust, but also for disorder strengths capable
of destroying the topological features, thus shedding light on the
robustness of TAI phases in the intermediate and strong disorder regimes
where perturbative methods are not reliable. Our findings are relevant
to understand Chern insulating phases in real systems where disorder
is unavoidable. In particular, the striking similarity between the
phase diagram we obtain for Anderson disorder and measurements of
differential drift velocity for the Haldane model realized with cold
fermionic atoms \citep{Jotzu2014} suggests that disorder could, at
least partially, account for the observed deviations from the result
expected for clean systems.

The paper is organized as follows: In Sec.\,\ref{sec:method}, we
introduce the Haldane model and specify the types of applied disorder.
We also briefly explain the methods employed to obtain the results
in Sec.\,\ref{sec:results}. In the latter, we present numerical
results on the phase diagram of the disordered Haldane model, including
the behavior of the topological phases up to the point they are suppressed
and the gapped and gapless regions of the phase diagram. Sec.\,\textcolor{blue}{\ref{sec:discuss}}
is dedicated to the discussion of the obtained results and in Sec.\,\ref{sec:conclusions}
the key results are summarized and some conclusions are drawn on their
implications.

\section{Model and methods}

\label{sec:method}

We consider the Haldane model \citep{Haldane1988} with Hamiltonian
written as
\begin{align}
H_{0}=-t\sum_{\left\langle i,j\right\rangle }c_{i}^{\dagger}c_{j}+t_{2}\sum_{\left\langle \left\langle i,j\right\rangle \right\rangle }e^{i\phi_{i,j}}c_{i}^{\dagger}c_{j}+\eta\sum_{i}\zeta_{i}c_{i}^{\dagger}c_{i},+\text{H.c}.,\label{eq:haldane_hamil}
\end{align}
where the disorder effects are induced by a site-dependent potential
term 

\begin{equation}
\begin{aligned}H=H_{0}+\sum_{i}\xi(i)c_{i}^{\dagger}c_{i}.\end{aligned}
\label{eq:disordered_haldane_hamil}
\end{equation}
Here, $c_{i}^{\dagger}(c_{i})$ are creation (annihilation) operators
defined in the two triangular sub-lattices $A$ and $B$ that form
the honeycomb lattice. The first term in Eq.\,(\ref{eq:haldane_hamil})
corresponds to hopping between nearest neighbor sites $\langle i,j\rangle$,
and couples sublattices $A$ and $B$. The second term describes complex
next-to-nearest neighbor hopping between sites $\langle\langle i,j\rangle\rangle$,
with amplitude $t_{2}e^{i\phi_{ij}}$ and $\phi_{ij}=\nu_{ij}\phi$,
where $\nu_{ij}=(2/\sqrt{3})(\hat{\boldsymbol{d}}_{1}\times\hat{\boldsymbol{d}}_{2})=\pm1$
with $\hat{\boldsymbol{d}}_{1}$ and $\hat{\boldsymbol{d}}_{2}$ two
unit vectors along the two bonds connecting $\langle\langle i,j\rangle\rangle$.
In the following $t$ is set to unity and $t_{2}=0.1t$ which ensures
a direct gap with no band overlapping \citep{Haldane1988}. The third
term corresponds to a staggered potential, where $\zeta_{i}=1$ if
$i\in A$ and $\zeta_{i}=-1$ if $i\in B$. These last two terms are
respectively responsible for breaking time-reversal and inversion
symmetries and thus for opening non-trivial and trivial topological
gaps at the Dirac points. In the absence of disorder the phase diagram
of the Haldane model in the $(\eta,\phi)$ parameter space encompasses
a trivial phase with vanishing Chern number and two topological non-trivial
phases with $C=\pm1$ respectively. The site-dependent potentials
$\xi(i)$ are uncorrelated for different sites and follow the probability
distribution

\begin{equation}
\begin{cases}
P_{W}(\xi_{i})=\frac{1}{W}\Theta\left(\frac{W}{2}-\left|\xi_{i}\right|\right) & \text{Anderson},\\
P_{V}(\xi_{i})=\frac{1}{2}\left[\delta\left(\xi_{i}\right)+\delta\left(\xi_{i}+V\right)\right] & \text{binary},
\end{cases}\label{eq:disorder_distrs}
\end{equation}
where $W$ and $V$ parametrize the disorder strength for the Anderson
and binary cases, respectively. 

In the following, we extend the phase diagram of the Haldane model
by increasing the disorder strength until the topological phases are
destroyed. We identify the topological nature of each phase by computing
the Chern number using Fukui's method \citep{Fukui2005} as implemented
in Ref.~\citep{Zhang2013}, a variant that is suitable to deal with
disordered systems for which translational invariance is broken. The
results are confirmed using the transfer matrix method (TMM) \citep{MK81,MacKinnon1983,hoffmann2012computational}
in the regions of the phase diagram where the density of states (DOS)
is gapless for which results are trustworthy. The method considers
a finite system with a fixed large longitudinal dimension $L$ and
a transverse dimension of size $M$, which is varied in order to compute
the localization length $\lambda_{M}$. We study the behavior of the
normalized localization length $\Lambda_{M}=\lambda_{M}/M$ as a function
of $M$: if $\Lambda_{M}$ decreases with $M$, the eigenstates are
localized in the thermodynamic limit and therefore the system is an
insulator; on the contrary, if $\Lambda_{M}$ increases with $M$,
the eigenstates are extended and the system is metallic; a constant
$\Lambda_{M}$ signals a critical point separating the two regimes.
The longitudinal dimension $L$ was chosen to be of the order of $10^{6}$
to guarantee a relative error smaller than $1\%$ for $\lambda_{M}$.
The DOS is obtained using a recursive Green's function method \citep{ehrenreich1980solid,Haydock1984},
allowing the access to system sizes in excess of $10^{6}$~lattice
sites.

\section{Results}

\label{sec:results}

\subsection{Phase diagram evolution}

\begin{figure*}
\begin{centering}
\includegraphics[width=0.98\textwidth]{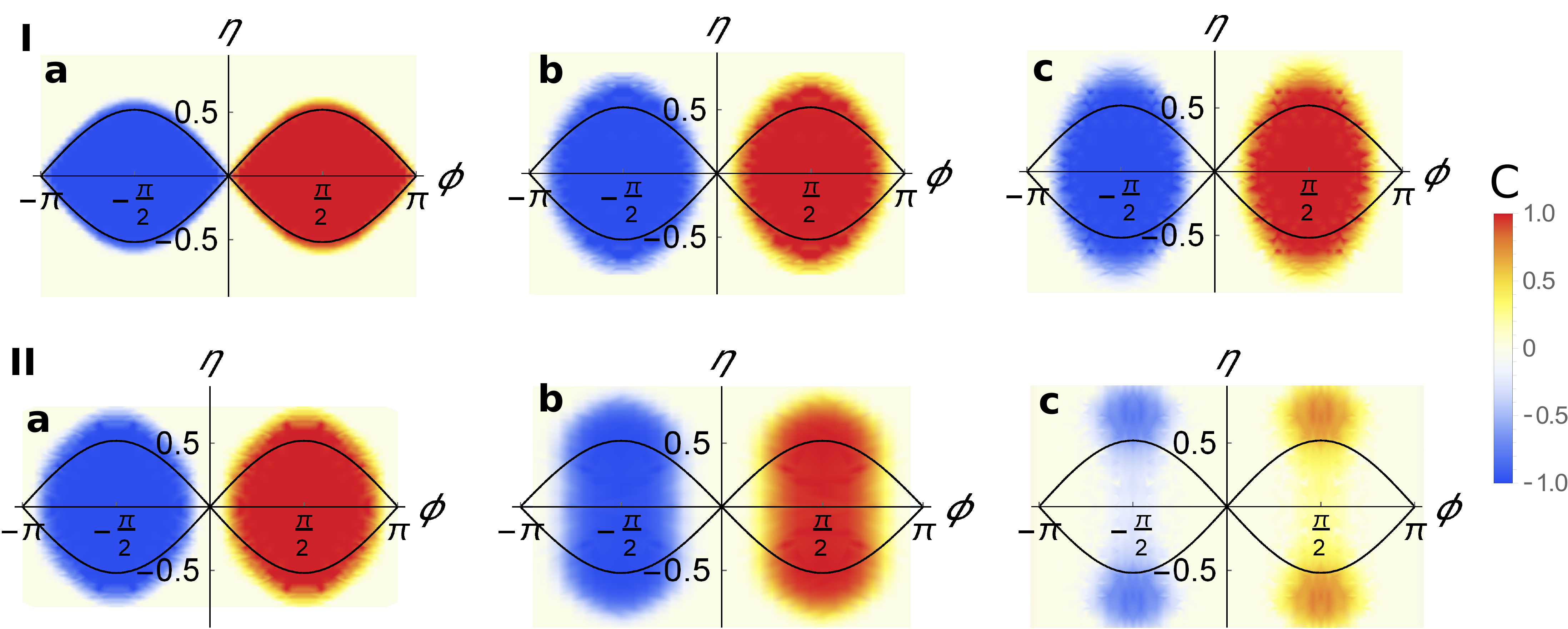}
\par\end{centering}
\caption{\label{fig:phase_diagrams_evo}(I) Evolution of the phase diagram
with Anderson disorder. a) $W/t=2$; b) $W/t=3.5$; c) $W/t=4$. The
results in a) and b) were obtained for a system of size $12\times12$
while the results in c) were for a $20\times20$ system. A total of
$100$ disorder configurations were used. (II) Evolution of the phase
diagram with binary disorder. a) $V/t=2$; b) $V/t=2.4$; c) $V/t=2.75$.
The results in a) were obtained for a $12\times12$ system while the
ones in figures b) and c) were obtained for systems of size $20\times20$.
The black curves correspond to the phase transition curves of the
Haldane model with no disorder.}
\end{figure*}

In this section we provide the evolution of the phase diagram with
the disorder strength for the Anderson and binary cases. Figure\,\ref{fig:phase_diagrams_evo}(I)
shows the evolution of the phase diagram for different disorder strengths
$W$ in the Anderson case. The color code depicts the values of the
Chern number and the black lines show the phase boundaries for the
case with no disorder. For small disorder strength, the topological
phases are robust and no significant change in the phase diagram is
observed. For $W/t=2$, a small enhancement of the topological phases
along the $\eta$ direction, shown in Fig.\,\ref{fig:phase_diagrams_evo}(Ia),
can already be observed for values of $\phi$ near $\pm\pi/2$. This
effect becomes very distinct for larger disorder strengths, as can
be seen in Figs.\,\ref{fig:phase_diagrams_evo}(Ib) and~\ref{fig:phase_diagrams_evo}(Ic)
and is in accordance with the results obtained in Ref.\,\citep{Sriluckshmy2018}.
Along with this phenomenon, the phases separate near $\phi=0$, becoming
``squeezed'' in the $\phi$ direction.

The evolution of the phase diagram for binary disorder is shown in
Fig.\,\ref{fig:phase_diagrams_evo}(II). The same qualitative phenomena
as for Anderson disorder is observed - an enhancement of the topological
phases in the $\eta$ direction and ``squeezing'' in the $\phi$
direction, as can be seen in Figs~\ref{fig:phase_diagrams_evo}(IIa)
and~\ref{fig:phase_diagrams_evo}(IIb). However, just before the
topological phases are destroyed, a different phenomenon occurs: the
last regions of the topological phases to disappear are for finite
$\eta$ {[}see Fig.\,\ref{fig:phase_diagrams_evo}(IIc){]}, in contrast
with the Anderson case for which this phenomenon does not occur. For
Anderson disorder, the squeezing shown in Figs.~\,\ref{fig:phase_diagrams_evo}(Ib)
and~\ref{fig:phase_diagrams_evo}(Ic) keeps going till the topological
phase is concentrated on a thin region around $\phi\simeq\pm\pi/2$,
including $\eta=0$, disappearing above a critical disorder $W_{c}/t\simeq5.2$.

\begin{figure}
\centering{}\includegraphics[width=0.7\columnwidth]{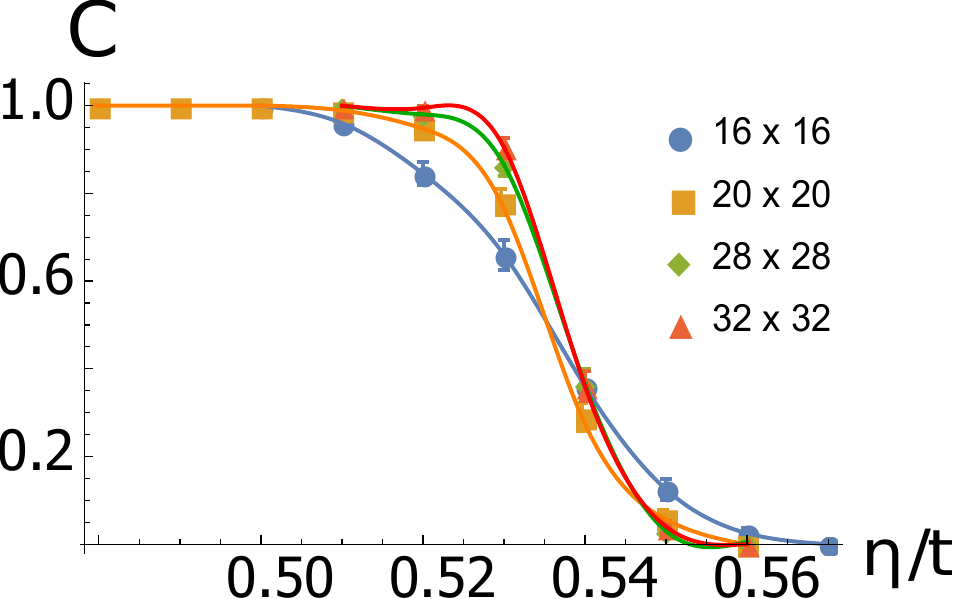}\caption{\label{fig:example_chern_curve}Example of the computed Chern numbers
for different system sizes for Anderson disorder with $W/t=1$ and
$\phi=\pi/2$. The standard deviation errors obtained by averaging
over different disorder realizations are shown for each data point.}
\end{figure}

To inspect the enhancement of the topological phases in the $\eta$
direction more quantitatively, we study the $(W,\eta)$ phase diagram
for $\phi=\pi/2$. We performed a finite size scaling analysis to
check the convergence of the phase transition point at the thermodynamic
limit. An average over 200 disorder configurations was always performed.
For each system size, the data points were interpolated. The phase
transition point was considered to be the intersection between the
curves corresponding to the larger systems. An example of the obtained
curves is provided in Fig.\,\ref{fig:example_chern_curve}. Notice
that the transition from $C=1$ to $C=0$ becomes sharper for larger
system sizes, precluding the abrupt transition in the thermodynamic
limit. The errors shown in Fig.~\ref{fig:phase_diagrams_evo} are
associated with the intersection of the two cubic splines used to
compute the phase transition points. Horizontal and vertical error
bars were respectively obtained by varying the disorder strength with
fixed $\eta$ and vice-versa.

\begin{figure*}
\begin{centering}
\includegraphics[width=0.98\textwidth]{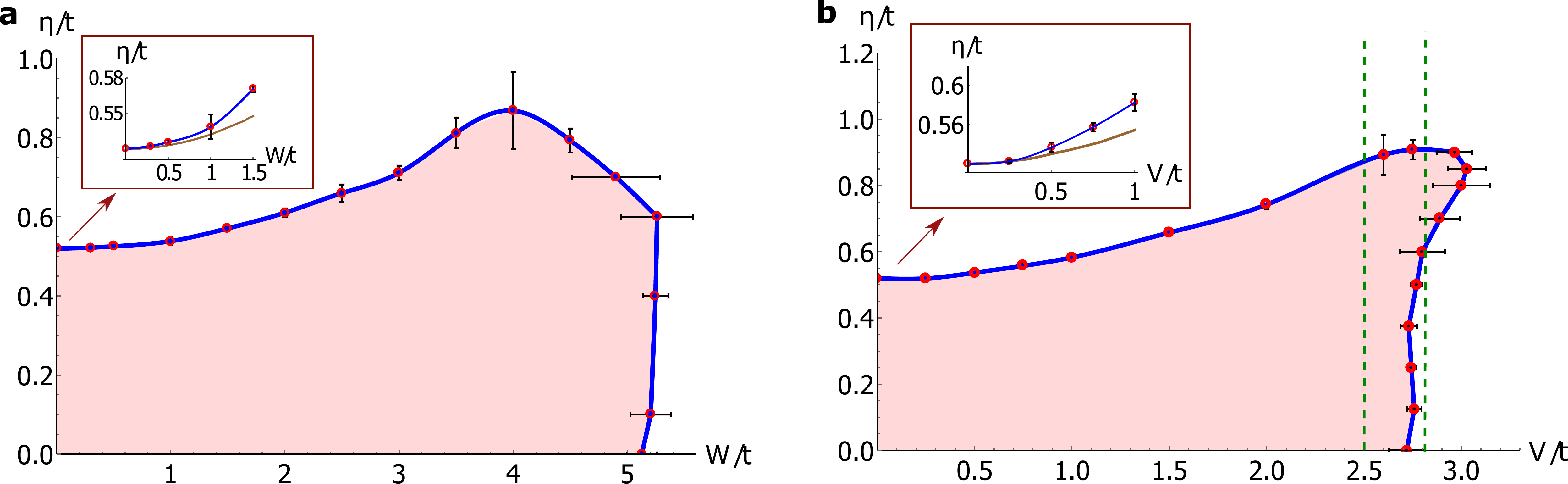}
\par\end{centering}
\centering{}\caption{\label{fig:phase_diagramDiseta}Phase diagram of the Haldane model
with Anderson disorder in the $(W,\eta)$ plane (a) and with binary
disorder in the $(V,\eta)$ plane (b), for $\phi=\pi/2$. The insets
show a region of small disorder amplitude zoomed to highlight the
comparison between the analytical predictions (brown curves), obtained
with a first-order self-consistent Born approximation (see Appendix~\ref{secap:self_cons_Born}),
and the numerical results. The dashed green lines in (b) correspond
to the cuts of the phase diagram analyzed with TMM in Fig.~\ref{fig:MTT_binary}. }
\end{figure*}

For Anderson disorder, the $(W,\eta)$ phase diagram at fixed $\phi=\pi/2$
is shown in Fig.\,\ref{fig:phase_diagramDiseta}(a) along with the
perturbative results shown in the inset, obtained for the first order
self-consistent Born approximation (see Appendix.~\ref{secap:self_cons_Born}).
The agreement between the perturbative result and the numerical calculation
in the low disorder regime is indicative of the correctness of the
later. For finite $W/t\lesssim5.1$, the transition from the trivial
insulating phase to the topological insulating one occurs for values
of $\eta$ larger then those attained for the case of no disorder
(where $\eta_{c}=3\sqrt{3}t_{2}$). This is a clear signature of the
onset of a TAI phase, which extends well beyond the validity of the
perturbative regime.

The $(V,\eta)$-phase diagram for binary disorder at fixed $\phi=\pi/2$
is shown in Fig.\,\ref{fig:phase_diagramDiseta}(b). The stabilization
of the topological phase for values of $V$ higher than in the clean
limit indicates the presence of a TAI phase also for binary disorder.
This agrees with the result from the first order self-consistent Born
approximation (see Appendix.~\ref{secap:self_cons_Born}), and shows
that, like for Anderson disorder, the TAI phase extends to intermediate
values of disorder, beyond the perturbative result. At odds with Anderson
disorder, however, a reentrant behavior is clearly seen in Fig.\,\ref{fig:phase_diagramDiseta}(b)
for binary disorder. This form of the $(V,\eta)$ transition line
confirms the phase diagram in the $(\phi,\eta)$-plane shown in Fig.\,\ref{fig:phase_diagrams_evo}(IIc),
where it is clearly seen that the topological region at $\eta$ close
to zero is destroyed by disorder effects before the regions at larger
values of $\eta$. 

\begin{figure}
\centering{}\includegraphics[width=0.98\columnwidth]{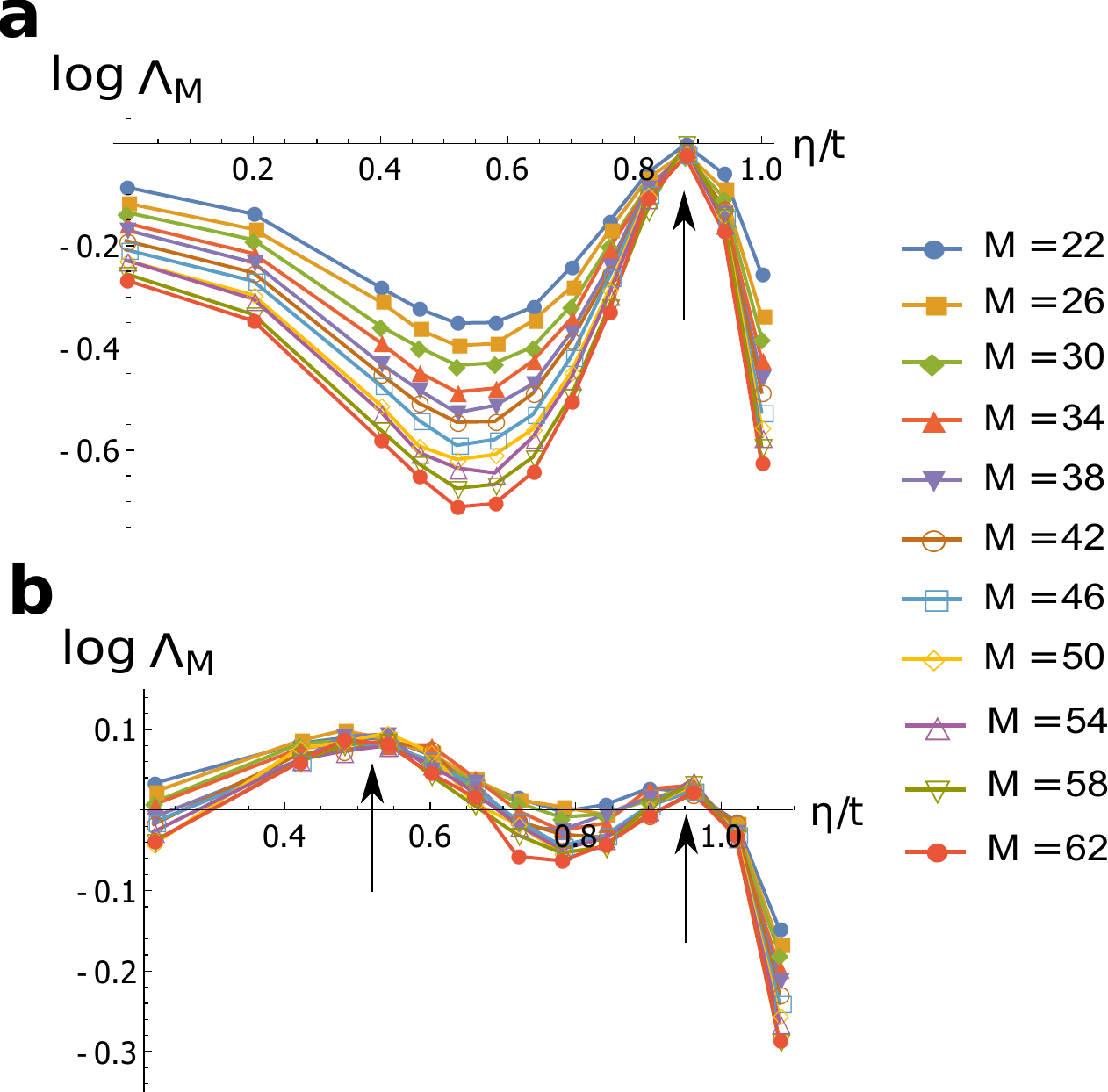}\caption{\label{fig:MTT_binary}Results obtained with TMM for binary disorder
$V/t=2.5$ (a) and $V/t=2.8$ (b) for systems with $M$-sites in the
transverse direction and periodic boundary conditions. The normalized
localization length $\Lambda_{M}$ decreases with $M$ except for
the phase transition points for which this quantity remains unchanged.
The single critical point obtained for $V/t=2.5$ and the two critical
points obtained for $V/t=2.8$ are in accordance with the results
obtained for the Chern number in Fig.\,\ref{fig:phase_diagramDiseta}(b).}
\end{figure}

\subsection{Localization properties}

The localization properties of disordered Chern insulators are well
understood \citep{OMN+03,nagaosaQSHloc07,sheng2012criticalMetal}.
This model belongs to class~A in the Altland-Zirnbauer symmetry classification
\citep{AZ97,evers-at2008}, the same as quantum Hall insulators, where
it is known that disorder localizes all states except those carrying
the Chern number (with known exceptions when spin rotation is broken
\citep{niu2016Metal,avishai2016Metal,avishai2015criticalMetal,sheng2012criticalMetal},
which is not the case here). As disorder increases, the bulk extended
states above and below the Fermi level carrying the topological invariant
shift toward one another and annihilate, leading to the topological
phase transition into the trivial phase through ``levitation and
annihilation'' of extended states \citep{nagaosaQSHloc07}. If we
fix the energy to the point where the two extended states annihilate
for some critical set of parameters $(W_{c},\eta_{c})$, the behavior
with increasing disorder of the normalized localization length $\Lambda_{M}=\lambda_{M}/M$,
obtained within the TMM, is as follows: for $W<W_{c}$ or $W>W_{c}$
the normalized localization length $\Lambda_{M}=\lambda_{M}/M$ decreases
with $M$, as expected for localized states, while right at the transition
point when $W=W_{c}$ the behavior $\Lambda_{M}=\text{const}$ is
expected, characteristic of an extended (critical) state. Exactly
the same behavior applies as a function of $\eta$ for fixed disorder.
We may therefore use the TMM to independently confirm the phase diagram
of Fig.~\ref{fig:phase_diagramDiseta} in the regions where the system
is gapless (see Sec.~\ref{subsec:gap}). 

For the disordered Haldane model studied here, it can be shown that
electron-hole symmetry is preserved at $\phi=\pi/2$. This guarantees
that the Fermi level at half-filling coincides precisely with the
energy where extended states meet and annihilate with increasing disorder
or trivial mass $\eta$. Therefore, if we plot $\Lambda_{M}$ at the
Fermi level as a function of $\eta$ the behavior described above
is expected, and the points where $\Lambda_{M}$ does not change with
$M$ should coincide with the phase transition lines in Fig.~\ref{fig:phase_diagramDiseta}
obtained through the analysis of the topological invariant. Particularly
relevant is the confirmation of the reentrant behavior found for binary
disorder, shown in Fig.~\ref{fig:phase_diagramDiseta}(b). Figure\,\ref{fig:MTT_binary}(a)
shows that for $V/t=2.5$ there is a single value of $\eta\simeq0.875$
for which $\Lambda_{M}$ becomes constant with increasing $M$ ranging
from $M=22$ to~$62$, signaling a phase transition. Away from the
critical point, $\Lambda_{M}$ decreases with $M$, implying that
the underlying phases are insulating. For $V/t=2.8$, Fig.~\ref{fig:MTT_binary}(b)
shows two transition points. Together, these findings confirm the
scenario shown in Fig.~\ref{fig:phase_diagramDiseta}(b) and further
validate the values of the critical points obtained by exact diagonalization.
We note that the quantum criticality of the Chern-to-trivial insulator
transition has been analyzed in Ref.~\citep{Prodan2013} for Anderson
disorder, and the same conclusions are expected to hold in the case
of binary disorder.

\subsection{Gapped and gapless regions of the phase diagrams}

\label{subsec:gap}

\begin{figure*}
\begin{centering}
\includegraphics[width=0.98\textwidth]{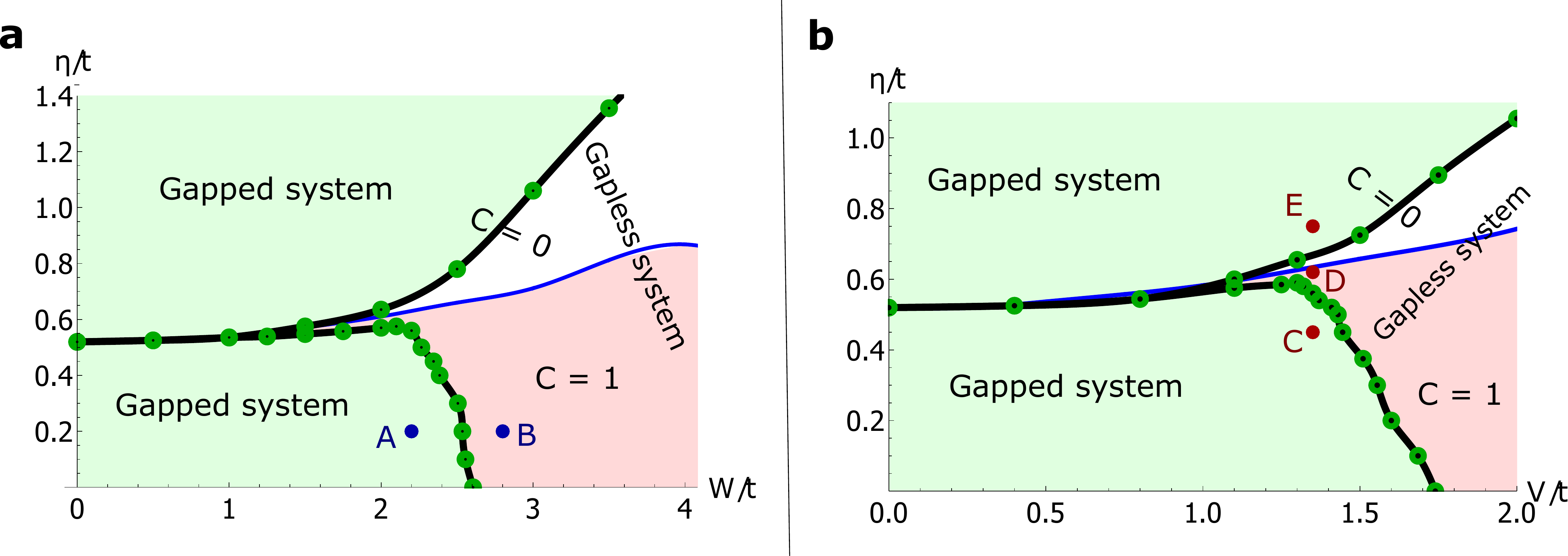}
\par\end{centering}
\caption{\label{fig:gapped_regions}Characterization of the system spectral
properties at the Fermi level in terms of its gapped or gapless nature
as a function of parameters $(W,\eta)$ for Anderson disorder (a)
and $(V,\eta)$ for binary disorder (b). We have fixed $\phi=\pi/2$.
Thick (black) lines with circles separate gapped from gapless regions,
while the thin (blue) line separates trivial from topological phases,
as in the phase diagrams of Fig.~\ref{fig:phase_diagramDiseta}.
The computations were carried out for systems of size $1000\times1000$.
The DOS for the points A-E marked in this figure is shown in Fig.~\ref{fig:DOS_points}.
The system was considered gapped whenever the DOS was below a threshold
value of $0.1\%\epsilon$, where $\epsilon$ was chosen to be $\epsilon=1/\Delta E$,
with $\Delta E=6t$ being the band width of the non-disordered system.
Point coordinates: A=(2.2,0.2); B=(2.8,0.2); C=(1.35,0.45); D=(1.35,0.62);
E=(1.35,0.75).}
\end{figure*}

We now turn to the gapped or gapless nature of the spectrum at the
Fermi level for the half-filled systems. In order to get this information,
we computed the DOS at the Fermi level using the methodology referred
in Sec.~\ref{sec:method}. Figure~\ref{fig:gapped_regions} shows
the phase diagram for $\phi=\pi/2$. The system was considered gapped
whenever the DOS was below a threshold value of $0.1\%\epsilon$,
where $\epsilon=1/\Delta E$ is a reference DOS value set to be the
inverse bandwidth $\Delta E=6t$ of the non-disordered system. A variation
of $\pm10\%$ in this criterion showed not to change significantly
the results. The obtained spectral characterization is qualitatively
the same for Anderson {[}Fig.~\ref{fig:gapped_regions}(a){]} and
binary {[}Fig.~\ref{fig:gapped_regions}(b){]} disorders. As expected,
for small disorder the system is always gapped except at the topological
transition line. This is seen in Fig.~\ref{fig:gapped_regions}(a)
and~\ref{fig:gapped_regions}(b) whenever the gapped-gapless transition
line (thick, black line with circles) coincides with the topological
transition line (thin, blue line). However, for larger disorder, both
trivial and topological regions become gapless and the topological
phase transition is no longer associated to a spectral gap closing
and re-opening. 

Figure~\ref{fig:gapped_regions} is one of our main results, showing
that topology and the presence or absence of a zero energy gap can
be used to unravel the rich phase diagram of the disordered Haldane
model. 

\begin{figure}
\centering{}\includegraphics[width=0.98\columnwidth]{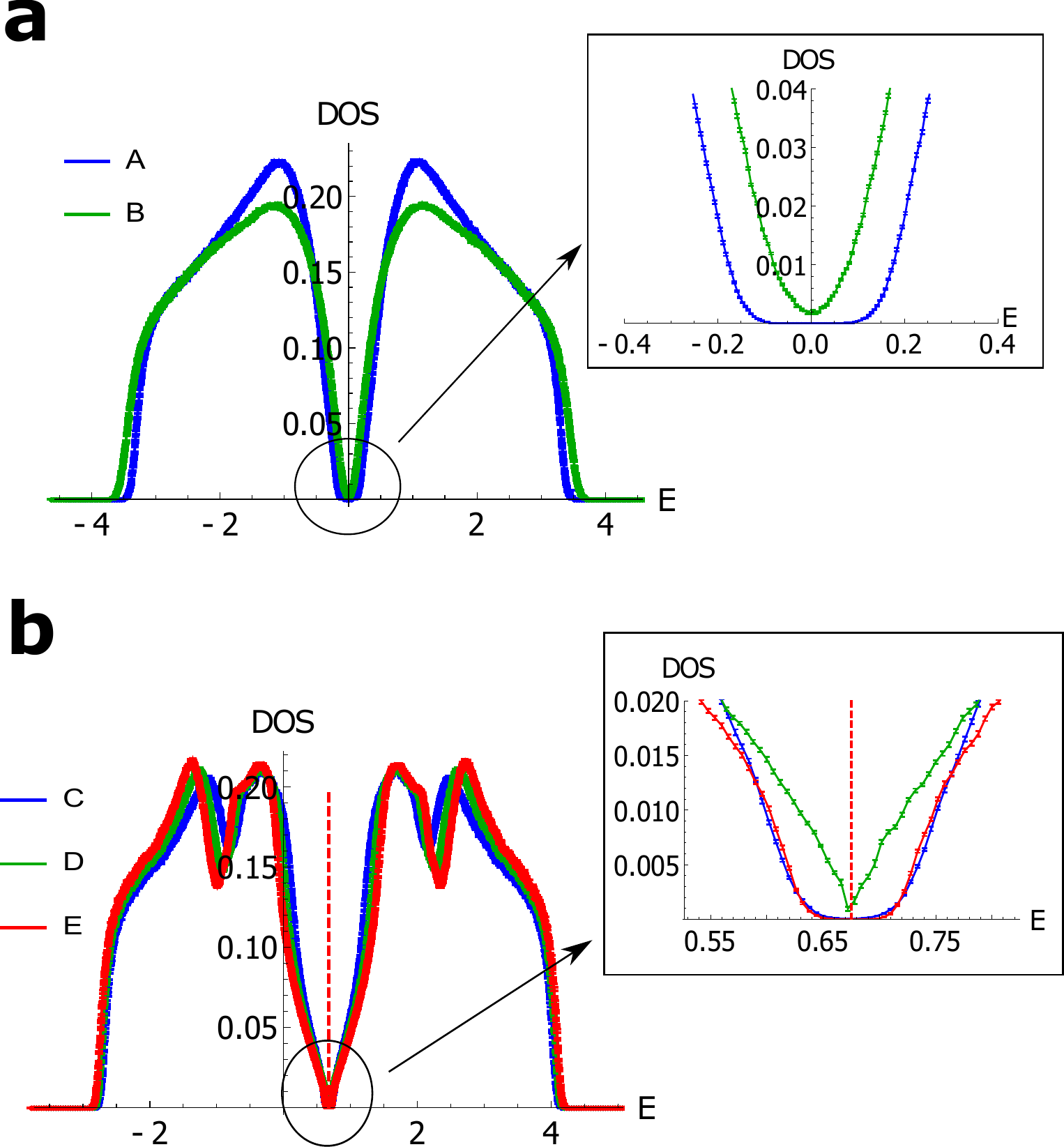}\caption{\label{fig:DOS_points}DOS for the points marked in Fig.~\ref{fig:gapped_regions}
for Anderson (a) and binary (b) disorder. The insets show a zoomed
region around $E=0$ for Anderson disorder and $V/2$ (corresponding
to the red dashed line) for binary, which corresponds to the Fermi
energy at half-filling in each case. An average over $25$ disorder
configurations was considered and the error bars correspond to the
standard deviation error associated to it. Systems of size $1000\times1000$
were used.}
\end{figure}

Examples of the DOS within the different regions of the phase diagram
are shown in Fig.~\ref{fig:DOS_points} for the points labeled A-E
in Fig.~\ref{fig:gapped_regions}. For Anderson disorder, close inspection
of the DOS at the Fermi level (zero energy) shows clearly {[}see inset
in Fig.~\ref{fig:DOS_points}(a){]} that in~A the system is gapped
while in B it is gapless. For binary disorder, where the Fermi level
at half-filling occurs at the energy $V/2$, it is seen that while
in points~C and~E the system is gapped {[}see inset in Fig.~\ref{fig:DOS_points}(b){]},
in point~D it is clearly gapless.

\section{Discussion}

\label{sec:discuss}

There are two notorious differences on how Anderson and binary disorder
affect the topological phases: (i) a larger degree of disorder is
necessary to destroy the topological phases in the Anderson case;
and (ii) in the case of binary disorder, the last regions of the topological
phases to be destroyed are for finite values of $\eta$, in contrast
with the Anderson disorder case, where a topological non-trivial region
around $\eta=0$ remains robust for higher disorder strength. These
results are corroborated by TMM calculations and a first order Born
approximation valid at small disorder strength. 

Regarding the difference in the critical disorder needed to destroy
the topological phase when we compare Anderson and binary disorder,
we note that such difference remains even if we make the comparison
in terms of the variance $\sigma^{2}$ of the respective disorder:
Anderson $\sigma^{2}=W^{2}/12$ and binary $\sigma^{2}=V^{2}/4$ (see
Appendix~\ref{secap:self_cons_Born}). Without better explanation,
we attribute this difference to the fact that Anderson disorder always
includes configurations of small local potential, thus small disorder,
as opposed to binary disorder for which the difference between site
potentials is always either $0$ or $V$.

Arguably, the most salient difference between the two types of disorder
is the reentrant behavior shown in Fig.~\ref{fig:phase_diagramDiseta}(b)
for binary disorder. Interestingly, if we fix $V\simeq2.8t$, the
reentrant topological phase shows up when the staggered potential
$\eta$ starts to be comparable with $V/2$. A possible explanation
for the effect would be the presence of some degree of cancellation
of the two potentials, the staggered potential and the disorder potential,
which would occur preferably for $\eta\sim V/2$.

The enhancement of the topological phases for moderate disorder strength
is a clear signature of TAI: regions of the phase diagram that are
trivial in a clean system are made topologically non-trivial by increasing
the disorder strength. This effect is very robust as it is clearly
seen for both types of disorder. For small disorder, an explanation
based on the renormalization of the gap, as given by the perturbative
treatment (see Appendix~\ref{secap:self_cons_Born}), provides a
reasonable understanding on the phenomenon. However, such explanation
does not work at higher values of disorder since, after the spectral
analysis shown in Fig.~\ref{fig:gapped_regions}, the TAI regions
become gapless for moderate disorder. More study is needed in order
to understand the origin of the robustness of the TAI phase in this
regime.

\section{Conclusions}

\label{sec:conclusions}

We have studied the Haldane model under the influence of Anderson
and binary disorder and obtained a comprehensive phase diagram for
increasing the disorder strength up to the point when no topological
phase survives. For both types of disorder, we obtained that the topological
phases are enlarged along the staggered lattice potential strength,
$\eta$, and ``squeezed'' along the flux $\phi$ direction. 

In addition to the topological classification, the presence or absence
of a spectral gap at the Fermi level was used to unravel the rich
phase diagram of the disordered Haldane model, which has been shown
to support four different phases: the gapped trivial and non-trivial
topological phases that can be adiabatically connected to those of
the clean system; and two gapless phases, either topologically trivial
or not, that arise between the gapped ones for any finite value of
disorder. 

The fact that our simple example of a disordered Haldane model at
half-filling already shows clear signatures of TAI phases at moderate
to high disorder values is encouraging. We expect our findings to
help guiding future attempts to realize disorder-induced topological
non-trivial materials, and also to help in the understanding of Chern
insulating phases in real systems where disorder is unavoidable. The
Haldane model has already been experimentally realized in systems
of ultracold fermions trapped in an optical lattice~\citep{Jotzu2014}
and more recently its realization as a topological insulator laser
was theoretically proposed~\citep{Harari2018}. Moreover, the high
controllability of system parameters in ultracold atoms~\citep{Roati2008}
make them a very interesting test bed for disorder-induced phenomena
and, in particular, to study how different kinds of disorder affect
the topological phase. Indeed, the striking similarity between the
phase diagram we obtained for Anderson disorder {[}Fig.~\ref{fig:gapped_regions}(c){]}
and measurements of the differential drift velocity for the Haldane
model realized in Ref.~\citep{Jotzu2014} suggests that disorder
could at least partially account for the observed deviations from
the expected result.
\begin{acknowledgments}
The authors acknowledge partial support from FCT-Portugal through
Grant No. UID/CTM/04540/2013. PR acknowledges support by FCT-Portugal
through the Investigador FCT contract IF/00347/2014. 
\end{acknowledgments}

\appendix

\section{Self-consistent Born approximation \label{secap:self_cons_Born}}

In this section we show that the enhancement of the topological phases
in the $\eta$ direction can be predicted perturbatively for small
disorder strength by using the first order self-consistent Born approximation.
This phenomenon is particularly noticeable for $\phi=\pi/2$, thus,
in following, we consider only this case for simplicity. The low energy
expansion of the Haldane Hamiltonian in momentum space around the
two Dirac points, labeled $\bm{K}_{+}$ and $\bm{K}_{-}$, can be
written as 

\begin{equation}
H_{0}(\bm{k})=v(\tau_{z}\sigma_{x}k_{x}+\sigma_{y}k_{y})+\Big[\eta-\Big(3\sqrt{3}t_{2}-\frac{9\sqrt{3}}{4}t_{2}k^{2}\Big)\tau_{z}\Big]\sigma_{z}\label{eq:haldanel_low_energy}
\end{equation}
where $v=3t/2$ is the Fermi velocity and $\bm{\sigma}$ and $\bm{\tau}$
are Pauli matrices that act respectively on the sub-lattice and the
Dirac point's pseudo-spin sub-spaces. As noticed in Ref.~\citep{Song2012},
this Hamiltonian can be mapped into the low energy effective Hamiltonian
of an HgTe quantum well, widely studied in the context of TAI \citep{Shen2009,Jiang2009,Groth2009}.
Following this set of works, we use the self-consistent Born approximation
for which the self-energy can be determined by the self-consistent
equation \citep{bruus2004many}:

\begin{equation}
\Sigma=\frac{3\sqrt{3}}{2}\Big(\frac{\sigma}{2\pi}\Big)^{2}\int_{\text{BZ}}\left\langle \Big[G_{0}^{-1}-\Sigma\Big]^{-1}\right\rangle \label{eq:self_cons_self_eneergy}
\end{equation}
where $G_{0}^{-1}=E-H_{0}$ is the Green's function of the unperturbed
Hamiltonian, $\left\langle ...\right\rangle $ denotes the disorder
average and $\sigma^{2}$ is the variance of the disorder distribution,

\begin{equation}
\sigma^{2}=\begin{cases}
W^{2}/12\,, & \text{Anderson disorder}\\
V^{2}/4\,, & \text{binary disorder}
\end{cases}.
\end{equation}
The pre-factor $3\sqrt{3}/2$ corresponds to the area of the unit
cell (in units of $a^{2}$, where $a$ is the lattice spacing).

Parameterizing the self-energy as $\Sigma=\Sigma_{\mathcal{I}}\mathcal{I}+\Sigma_{x}\sigma_{x}+\Sigma_{y}\sigma_{y}+\Sigma_{z}\sigma_{z}$,
we can see that setting $\Sigma_{x}=\Sigma_{y}=0$ solves Eq.~\eqref{eq:self_cons_self_eneergy}
since the off-diagonal entries of $G_{0}$, which are proportional
to $k_{x}$ or $k_{y}$, vanish after integration. Since the integration
procedure yields a $\bm{k}$-independent $\bm{\Sigma}$, $\Sigma_{\mathcal{I}}$
and $\Sigma_{z}$ can be respectively seen as a renormalization of
the $\bm{k}$ independent terms of $G_{0}^{-1}$. Labeling the usual
Haldane's topological mass as $m=\eta-3\sqrt{3}t_{2}\tau_{z}$ , we
can write:

\begin{equation}
\begin{array}{cc}
m'=m-\Sigma_{z}\texttt{ }\texttt{ } & E'=E+\Sigma_{\mathcal{I}}\end{array}
\end{equation}
with $m'$ being the renormalized topological mass and $E'$ the energy
shift. Since we are imposing half-filling, the energy shift will be
absorbed by a change of the chemical potential and the Fermi level
will remain unchanged. The Chern number in the presence of disorder
can thus be given in terms of the difference between the renormalized
topological masses on Dirac points $\bm{K}_{+}$ and $\bm{K}_{-}$
as

\begin{equation}
C=\frac{1}{2}[sgn(m'_{+})-sgn(m'_{-})].
\end{equation}
This expression, was used to obtain the analytical phase transition
curves in Fig.~\ref{fig:phase_diagramDiseta}. The enhancement of
the topological phases in $\eta$ for small disorder strength is already
predicted within this simple approximation. For higher disorder strength,
however, the growth rate of the topological phase obtained numerically
becomes significantly larger than the one obtained within the Born
approximation both for Anderson and binary disorders. 

\bibliographystyle{apsrev4-1}
\bibliography{mybib}

\end{document}